\documentclass[aps,pre,twocolumn,showpacs,floatfix]{revtex4}

\usepackage{graphicx}
\usepackage{graphics}
\usepackage{latexsym}
\usepackage{subfigure}
\usepackage{multirow}
\usepackage{amsmath}
\usepackage{amssymb}
\usepackage{txfonts}
\usepackage[usenames]{color}

\newcommand{\be}{\begin{equation}}
\newcommand{\ee}{\end{equation}}
\newcommand{\bea}{\begin{eqnarray}}
\newcommand{\eea}{\end{eqnarray}}

\begin{document}

\title{Phase and frequency entrainment in locally coupled phase oscillators with repulsive interactions}

\author{Michael Giver, Zahera Jabeen and Bulbul Chakraborty}
\affiliation{Martin A. Fisher School of Physics, Brandeis University, Waltham, MA USA}

\date{\today}

\begin{abstract}
Recent experiments in one and two-dimensional microfluidic arrays of droplets 
containing Belousov 
-Zhabotinsky reactants  show a rich variety of spatial patterns [J. Phys. Chem. Lett.  1, 1241-1246 (2010)]. The dominant coupling between these droplets is inhibitory.
Motivated  by this experimental system, we study 
 repulsively coupled Kuramoto oscillators with nearest neighbor interactions, on a linear chain as well as a ring in one dimension, and on a triangular lattice in two dimensions.
In one dimension, we show using linear stability analysis as well as numerical study, that the stable phase patterns depend on the geometry of the lattice. We show that a transition to the ordered state does not exist in the thermodynamic limit. 
In two dimensions,
we show that the geometry of the lattice 
constrains the phase difference between two neighboring oscillators to $2\pi/3$. 
We report the existence of  
domains with either clockwise or anti-clockwise helicity,
 leading to defects in the lattice. 
We study the time dependence of these domains and show that at large coupling strengths the domains freeze due to frequency synchronization. Signatures of the above phenomena can be seen in the spatial correlation functions. 

\end{abstract}
 
\maketitle

\section{Introduction}

Synchronization~\cite{Strogatzbook,Strogatz:1993bs}, in which individual oscillators collectively 
organize into some phase and frequency relationship with each other is 
prevalent in nature.  It has been studied extensively in many biological and chemical systems. Examples of systems which exhibit partial or complete synchronization include  oscillations in neuronal networks \cite{heartcells,Skinner1994,Lisman1995,Jensen1996a,Jensen1996b}, and coupled chemical oscillators
\cite{Kiss:2002fk,Lengyel1991,Yang2000,Yang2003}. 

The simplest model that exhibits synchronization was proposed by Kuramoto \cite{kuramotobook},  and applies to weakly-coupled systems where the phase varies slowly compared to the intrinsic frequencies.  This is an exactly solvable model.  Even though the model is mean-field (every oscillator coupled to every other oscillator), and ignores amplitude variations \cite{Hong:2004dw}, it has been very successful in capturing the qualitative features of synchronization.  Locally coupled versions of the Kuramoto model have been studied using numerical techniques, and by mapping to statistical mechanics models \cite{Hong:2004dw,Hong:2005ss,Wood:2007qc,Wood:2006uf,Wood:2007jo,Wood:2006mw}.

Most of the well-studied models invoke attractive (or excitatory) coupling between the oscillators.  In these systems, a transition from ordered to disordered phases is seen at finite coupling strengths, and the lower critical dimension for frequency entrainment was shown to be $d=2$  \cite{Hong:2004dw,Hong:2005ss,Wood:2007qc,Ostborn:2009qf}.  
More recently, oscillators with repulsive (or inhibitory) couplings have generated interest due to applications to biological systems \cite{Campbell:1999ez,Balazsi:2001uq,Leyva:2006zr,Zhou:2008vn}.
Unlike the attractive coupling case, where all oscillators relax into an in phase state irrespective of the geometry, the solutions in the case of oscillators with inhibitory coupling depend strongly on the underlying geometry, since the connectivity of the lattice can frustrate the local order preferred by the coupling. In a one dimensional chain, the nature of the global pattern was seen to change when the coupling was changed from a local to a global coupling \cite{Tsimring:2005rq}.  The global inhibitory coupling is ``frustrated'' since every oscillator prefers to be $\pi$ out of phase with every other oscillator.  Other variants of repulsively coupled systems have also been studied \cite{moon_phaseshifted}. Synchronous, traveling waves which undergo a transition to spatiotemporal chaos were seen in coupled circle maps with repulsive coupling \cite{gadesinha}. 
The role of frustration in determining patterns and tuning adaptive networks has been studied in the context of biological systems~\cite{Krishna:2009fu,Daido:1992dz,Inoue:2010vn,Tiberkevich:2009ys}.

Recently, 
an experimental system with inhibitory coupling, and controllable geometry and coupling strength, was realized in 
an array of water microdroplets surrounded by an oil medium \cite{Toiya:2008fk,Vanag:2010uq,Toiya:2010kx}. 
These water droplets contained reactants of the oscillatory Belousov-Zhabotinsky (BZ) reaction. 
Bromine, which is a constituent reactant,  dissolves preferably in the oil medium, and provides the inhibitory coupling between the droplets.
The strength of the coupling between the droplets is varied by varying the size of the droplets.
In one dimensional array of these droplets, anti-phase synchronization
as well as Turing patterns are observed.
In two dimension geometry, these droplets rearrange into a hexagonal array.
At weak coupling strengths, the droplets relax into  a "$2\pi/3$" 
pattern in which any two neighboring droplets
 are $120^{\circ}$  out of phase. 
At large coupling strengths, a "$\pi-$S" pattern is seen, in which the central droplet relaxes into a non-oscillatory stationary state, and the surrounding six droplets  in the hexagonal array
exhibit anti-phase oscillations. This experiment was qualitatively explained by including diffusion of inhibitory species in the FKN model generally used to describe BZ reaction \cite{Toiya:2010kx}.

Motivated by this experimental system, we study a local variant of the Kuramoto model with repulsive coupling. We show that in one dimension (1D), the neighboring  oscillators prefer to relax into an anti-phase state. The  spatial pattern of  phase differences change when the geometry is changed from a linear chain to a ring.  For the 1D ring, the phase pattern depends on the number of oscillators.  We explain the observed patterns by studying attractors at infinite coupling using linear stability analysis.
We show that there is no synchronization transition  in the thermodynamic limit since the critical coupling strength scales with the number of oscillators.  In two dimensions (2D),  we study oscillators on a triangular lattice, where we expect the effects of frustration to be the most pronounced.
We show that the oscillators prefer to relax to the $2\pi/3$ state seen in the experimental system. 
We report the existence of  
domains with either clockwise or anti-clockwise helicity,
in which phases of any three neighboring oscillators 
either increase or decrease in a given direction,  leading to {domain-wall defects} in the lattice. 
We study the time dependence of these domains and show that at large coupling strengths 
the domains freeze into a ``glassy state'' in which the phases are disordered but the frequencies are synchronized. 
We characterize these phenomena by studying spatial correlation functions. 
Finally, we discuss these results in the context of the experimental BZ microdroplet system. 

The paper is organized as follows. We give details of the model in section \ref{sec1}. We discuss the results from our 1D study in section \ref{sec2}.
In section \ref{sec3}, we discuss the results obtained in 2D. We conclude with a discussion of our results in the context of the experimental system.
\section{The model \label{sec1}}

The BZ micro-oscillators were modeled using locally coupled Kuramoto oscillators placed on a lattice.  The equations governing the phase of the oscillators are: 
\begin{equation}
\dot{\phi}_i = \omega_i + K\sum_{\langle i j\rangle} \sin(\phi_i-\phi_j)
\label{phase-eqn}
\end{equation}
Here, the oscillator at site $i$ is coupled locally to its nearest neighbors $\lbrace j \rbrace$. The intrinsic frequency of the individual oscillators is given by $\omega_i$. This frequency is chosen randomly from a Gaussian distribution with zero mean and variance $\sigma$, and is a source of quenched disorder in this system \cite{Hong:2004dw}.  The strength of the coupling between the sites is given by $K$.  As discussed below, the effective coupling strength that controls the synchronization behavior is $K/\sigma$, and therefore, we set $\sigma$ to unity without loss of generality. When the coupling strength in the above system is attractive, viz. $K<0$, the oscillators synchronize in phase with each other. However, when $K>0$, the coupling is repulsive and resulting attractors depend intrinsically on the underlying geometry of the lattice. 

In 1D, we study a linear chain, as well as a ring geometry. In 2D, we study a triangular lattice, with six nearest neighbor connections. 
Both in 1D and 2D, the linear size of the array is $L=64$ unless otherwise specified, with the lattice spacing taken to be unity. The numerical integrations are performed using  Runge-Kutta algorithm with a fixed step-size $0.05$. We have checked the step-size dependence of our results, and except in the glassy state, which will be discussed in detail below, the results are not sensitive to reducing the step size beyond this value.  Statistical properties of the data are obtained by averaging over $100$ realizations of quenched frequencies in 1D and over $50$ realizations  in 2D.

\section{Results in 1D \label{sec2}}

\subsection{Linear Stability}
In 1D we consider the system with two different boundary conditions: free ends (linear chain), and periodic (ring). With the free end boundary conditions, the oscillators at either end of the array have only one neighbor to which they are coupled, whereas the oscillators form a ring in the case of periodic boundary conditions, in which the first oscillator is coupled to the $N^{th}$. As we will show, this can create frustration in the system, and  give rise to interesting spatial patterns when the oscillators are repulsively coupled. 

We begin our study in 1D by performing a linear stability analysis of the deterministic version of Eq. \ref{phase-eqn}.  In this system, there is no disorder in the  intrinsic frequencies: $\sigma=0$, and all the oscillators have the same frequency, $\omega$.  Since the effective coupling strength controlling phase and frequency patterns is $K/\sigma$, taking the limit of $\sigma \rightarrow 0$ is, therefore, equivalent to taking $|K| \rightarrow \infty$.  The linear stability analysis is thus a study of the stability of the attractors in the infinite coupling strength limit.

The linear stability analysis is most transparent if we perform a change of variable to the phase difference $\theta_i = \phi_i-\phi_{i+1}$ between nearest neighbors.  The transformed equations in 1D are:
\begin{equation}
\dot{\theta}_i = -K\sin(\theta_{i-1})+2K\sin(\theta_i)-K\sin(\theta_{i+1}).
\label{theta-phase-eqn}
\end{equation}
It is known \cite{ermentrout:215} that in a linear chain, Eqn.\ref{theta-phase-eqn} has $2^{N-1}$ attractors, but only one stable solution which depends on the sign of $K$. For $K<0$ only the completely in-phase solution, $\theta_i = 0$, is stable, while for $K>0$ only the completely anti-phase solution, $\theta_i = \pi$ is stable.

\begin{figure}
  \begin{center}

 \includegraphics[width=\columnwidth]{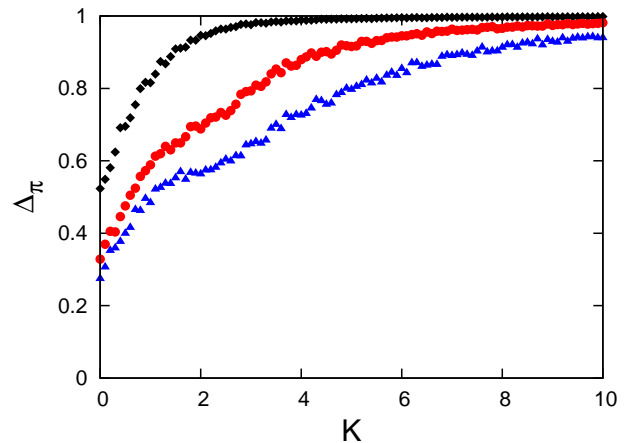}
  \caption{(Color online) Order parameter, $\Delta_\pi$, for $N = 3$($\Diamond$), $N = 7$($\bigcirc$), and $N = 10$($\triangle$) for linear chain averaged over random initial phase configurations.  }
  \label{fig-op-linearchain}
  \end{center}
\end{figure}
It is under periodic boundary conditions that we find a much richer long time behavior. Solving for the fixed points, we find the condition
\begin{equation}
\sin(\theta^\star_i) = \sin(\theta^\star_j)
\label{fixedpoints-eqn}
\end{equation}
for all $i,j$. Additionally, we have the constraint imposed by the boundaries
\begin{equation}
 \sum_{i = 1}^N\theta_i = 2\pi n
\end{equation}
where $n$ is an integer. Thus the system has fixed point solutions for all values $\theta^\star = 2\pi n/N$. We can determine the stability of these solutions by noting that the Jacobian is a circulant matrix \cite{circulant}, for which, the eigenvalues are easily computable. Specifically, we find
\begin{equation}
\lambda_m = 4 K\sin^2(m\pi/N)\cos\theta^{\star};~~m= 0,1,...,N-1.
\end{equation}
Since $\sin^2(m\pi/n)$ is always positive, the stability depends only on the phase difference $\theta^{\star}$, and the sign of $K$. For values of $K>0$, all values of $\theta^\star$ between $\pi/2$ and $3\pi/2$ are stable, where as for $K<0$, $\theta^\star$ between $-\pi/2$ and $\pi/2$ are stable.  It should be noted that while $\theta^\star = \pi$ is the most stable phase configuration for $K>0$, it is not accessible for systems with odd numbers of oscillators. Contrast this with the case of $K < 0$ where the most stable state, $\theta^\star = 0$, is accessible for any number of oscillators.  

We studied the synchronization transition in 1D by numerical integration of the dynamical equations for the  phase (Eq  .\ref{theta-phase-eqn}). Here, we take the intrinsic frequencies of our oscillators to be chosen randomly from a Gaussian distribution.The mean of the Gaussian is often set to zero, but the behavior of the system is the same for any value of the mean. There are two distinct processes leading to the complete synchronization of the system: frequency entrainment, where the oscillators adjust their frequency of oscillation until all evolve with the same frequency, and phase ordering, where the oscillators develop a uniform phase difference between all neighboring elements. {One interesting question that we address is whether the frequency and phase synchronization occur as two separate transitions, or is there a single transition where both phases and frequencies lock in.

\begin{figure}
  \begin{center}
 \includegraphics[width=\columnwidth]{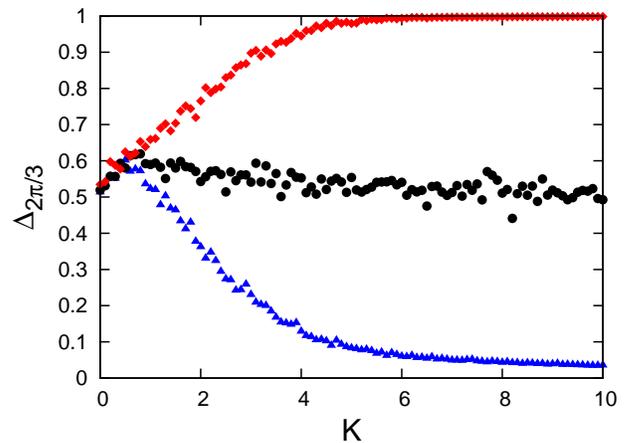} 
  \caption{(Color online) Order parameter, $\Delta_{2\pi/3}$, for $N=3$ on a ring. Three different initial phase configurations are shown: all initially in-phase($\bigcirc$), all initially $-2\pi/3$ out of phase($\triangle$), and initially $\pi/2$ out of phase($\Diamond$). }
  \label{fig-op-periodic}
  \end{center}
\end{figure}

\begin{figure}
  \begin{center}
\includegraphics[width=\columnwidth]{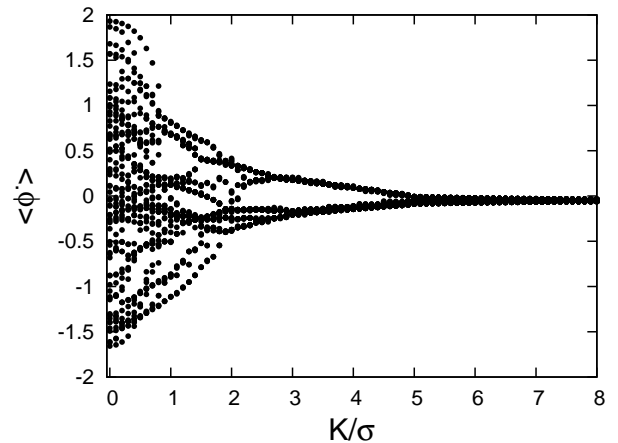}
  \caption{Bifurcation diagram of time averaged oscillator frequencies $\dot{\phi_i}$ plotted as a function of coupling strength $K$ for a ring of $64$ oscillators.  }
  \label{fig-synchtree}
  \end{center}
\end{figure}
\subsection{Phase Ordering}
To investigate the phase ordering we define a phase order parameter \cite{Hong:2005ss}, as
\begin{equation}
\Delta_q = \langle \frac{1}{N}|\sum_{j = 1}^N e^{i(\phi_j-(j-1)q)}|\rangle
\label{eq-orderparameter}
\end{equation}
which is unity for the fully ordered system, and zero when the system is completely disordered. Here, $q$ is the expected phase difference between oscillators and the average is over realizations of $\omega_i$.

As discussed above, we expect the system with free end boundary conditions to always go to the completely anti-phase state at high enough coupling strength. In Fig. \ref{fig-op-linearchain} the order parameter, $\Delta_{\pi}$ is plotted as a function of coupling strength for different system sizes. Although the initial phase configurations were chosen at random, $\Delta_{\pi}$ still tends to a value of one with increasing $K$, showing that there is only one attractor, $\theta^\star = \pi$. Additionally, we can see that as $N$ is increased a larger coupling is required to reach the completely ordered state. This agrees with the conclusions reached in Ref. \cite{Hong:2005ss} regarding attractively coupled oscillators, that that there can be no complete phase synchronization in the thermodynamic limit for $d<5$ \cite{Hong:2005ss}.

With periodic boundary conditions, the story becomes a bit different.  Fig. \ref{fig-op-periodic} shows the order parameter, $\Delta_{2\pi/3}$, for a three oscillator ring. The linear stability analysis of this system gives two stable states, $\theta^\star = \pm 2\pi/3$. We notice in the figure that the behavior of the order parameter depends upon the initial phases of the oscillators. If all oscillators are given the same initial phase, $\theta_0=0$ where $\theta_0 = \phi_i-\phi_{i-1}$, they are equally likely to fall into either of the $\pm2\pi/3$ attractors, thus $\Delta_{2\pi/3}$ takes on a value of roughly one half. If on the other hand, $0 < \theta_0 < \pi$ the system always goes to the $2\pi/3$ state and $\Delta_{2\pi/3}$ tends to one, while for $\pi < \theta_0 < 2\pi$, the system goes to the $-2\pi/3$ state and $\Delta_{2\pi/3}$ tends to zero.  These two states  in the three-oscillator ring appear in the 2D, triangular lattice as the two helicities of the $2\pi/3$ state, and lead to the frozen domains, as discussed in detail in section \ref{sec3}.

\begin{figure}
  \begin{center}
 \includegraphics[width=\columnwidth]{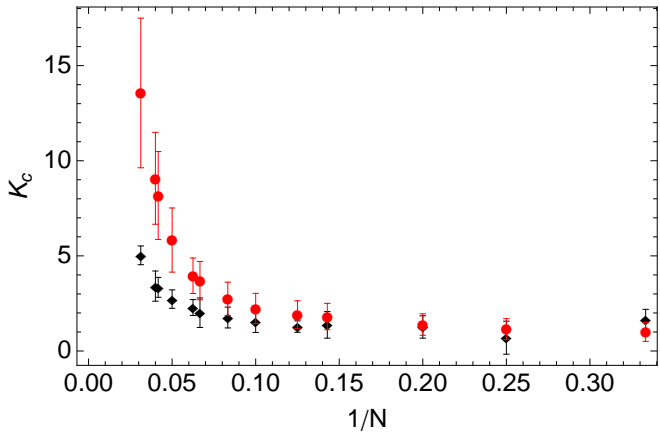}
  \caption{(Color online) Critical coupling strength as a function of $1/N$ where $N$ is the number of oscillators, for linear chain($\bigcirc$) and ring($\Diamond$) geometries.}
  \label{fig-critcoupling}
  \end{center}
\end{figure}


\begin{figure}
\begin{center}
	\begin{tabular}{cc}
		\includegraphics[width=5.cm,height=4.2cm]{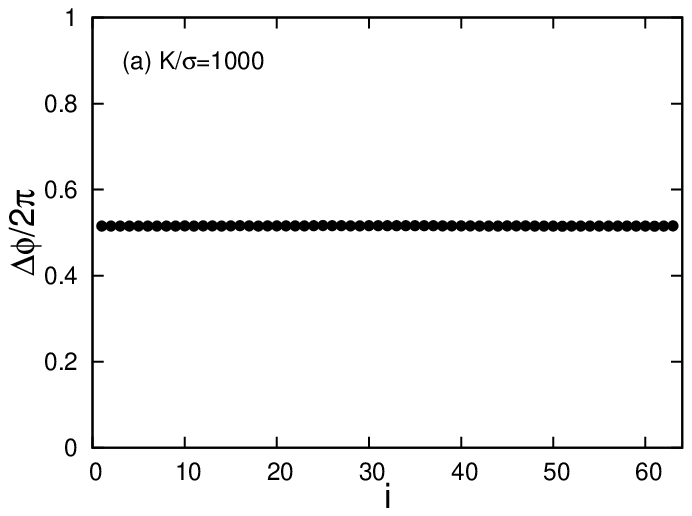}&
		\includegraphics[width=4.cm,height=4.0cm]{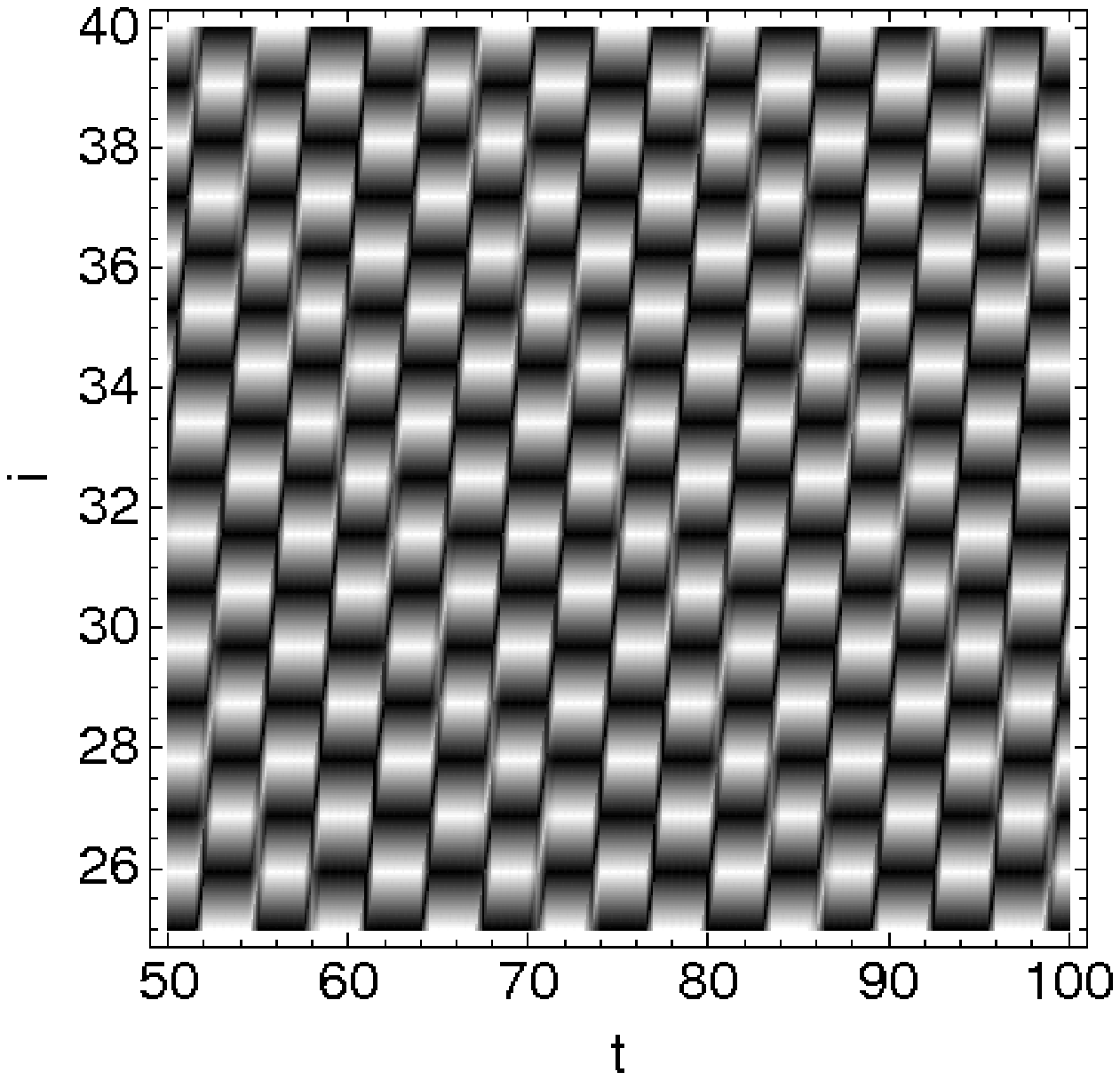}\\
		\includegraphics[width=5.cm,height=4.2cm]{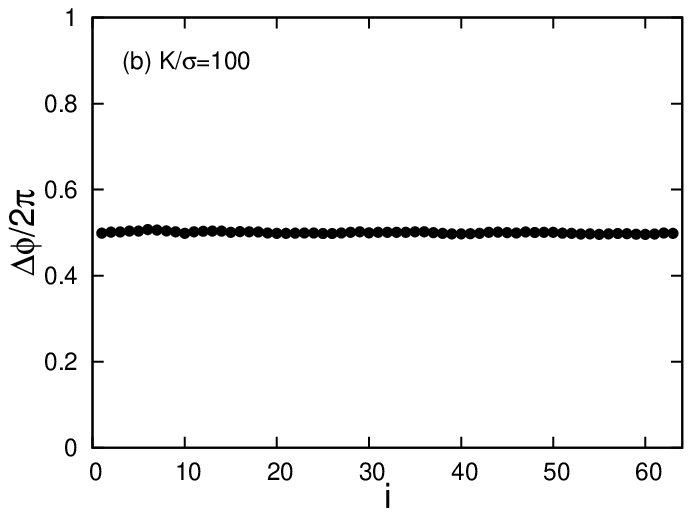}&
		\includegraphics[width=4.cm,height=4.cm]{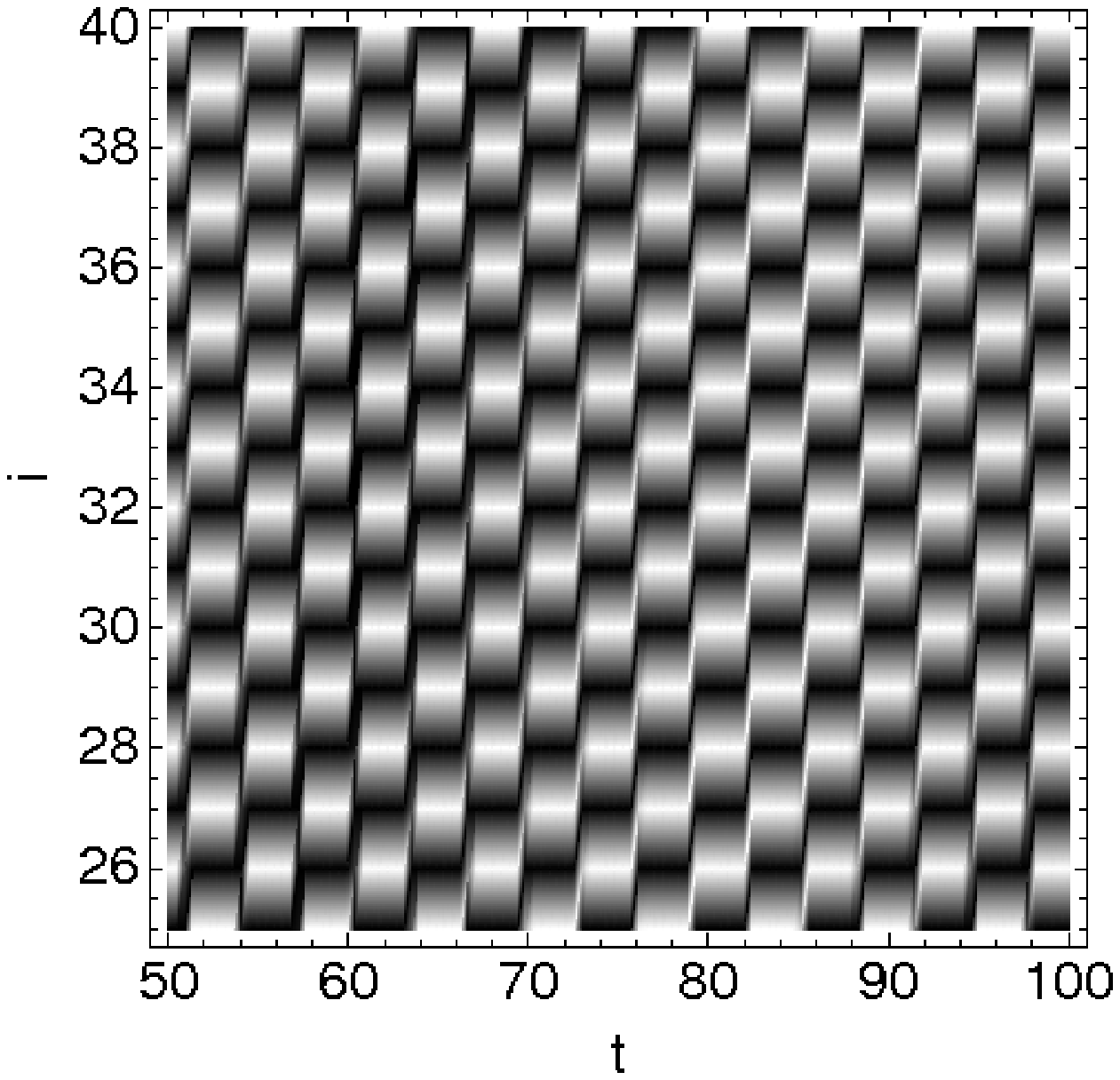}\\
\includegraphics[width=5.cm,height=4.2cm]{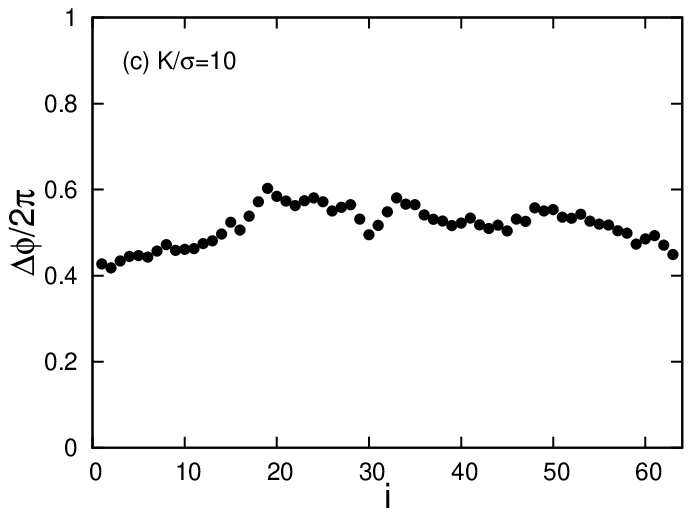}&
		\includegraphics[width=4.cm,height=4.cm]{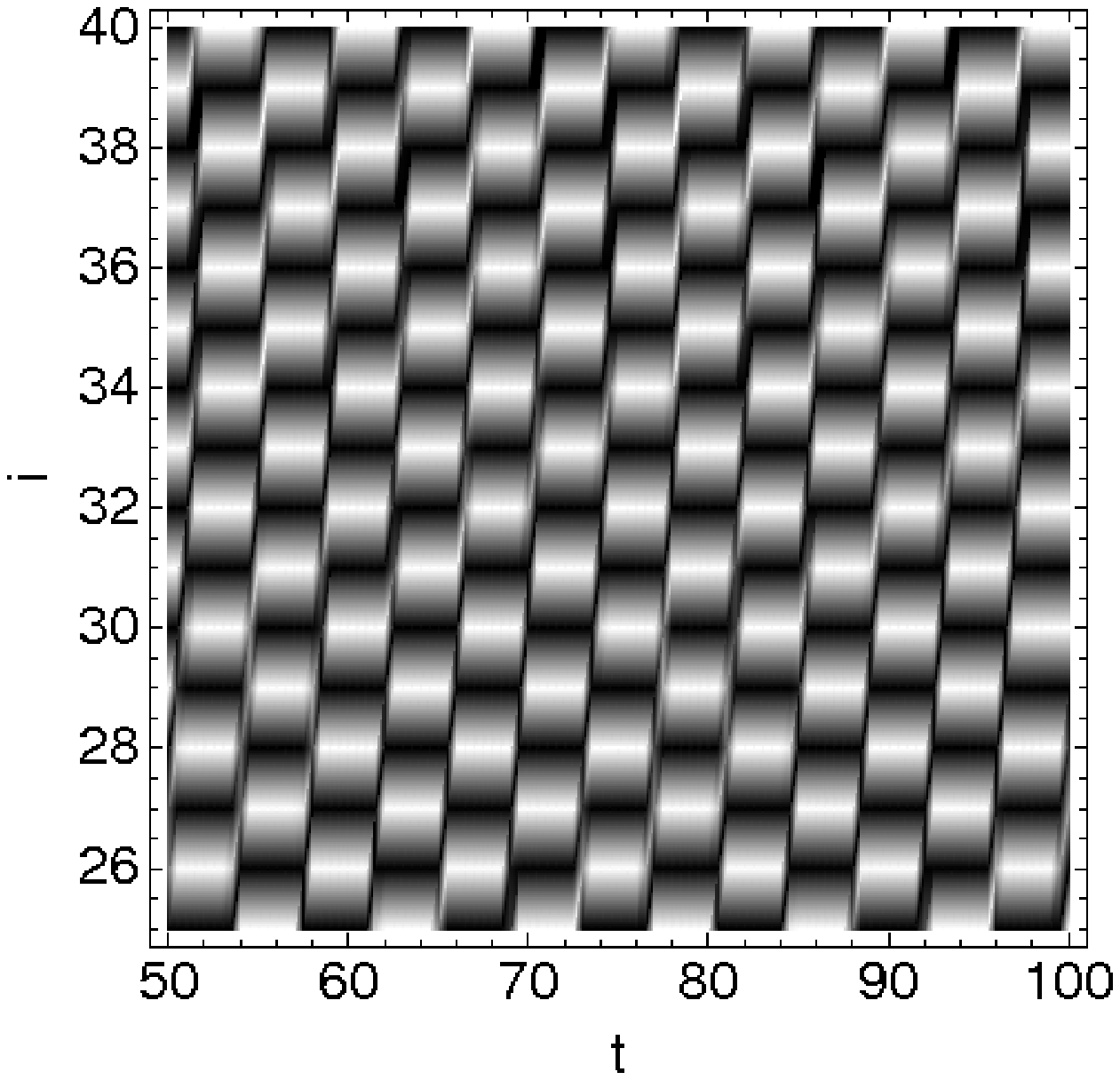}\\
		\includegraphics[width=5.cm,height=4.2cm]{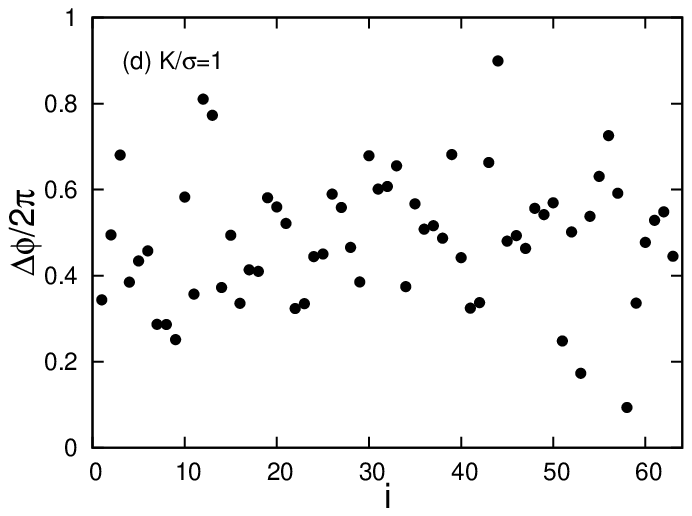}&
		\includegraphics[width=4.cm,height=4.cm]{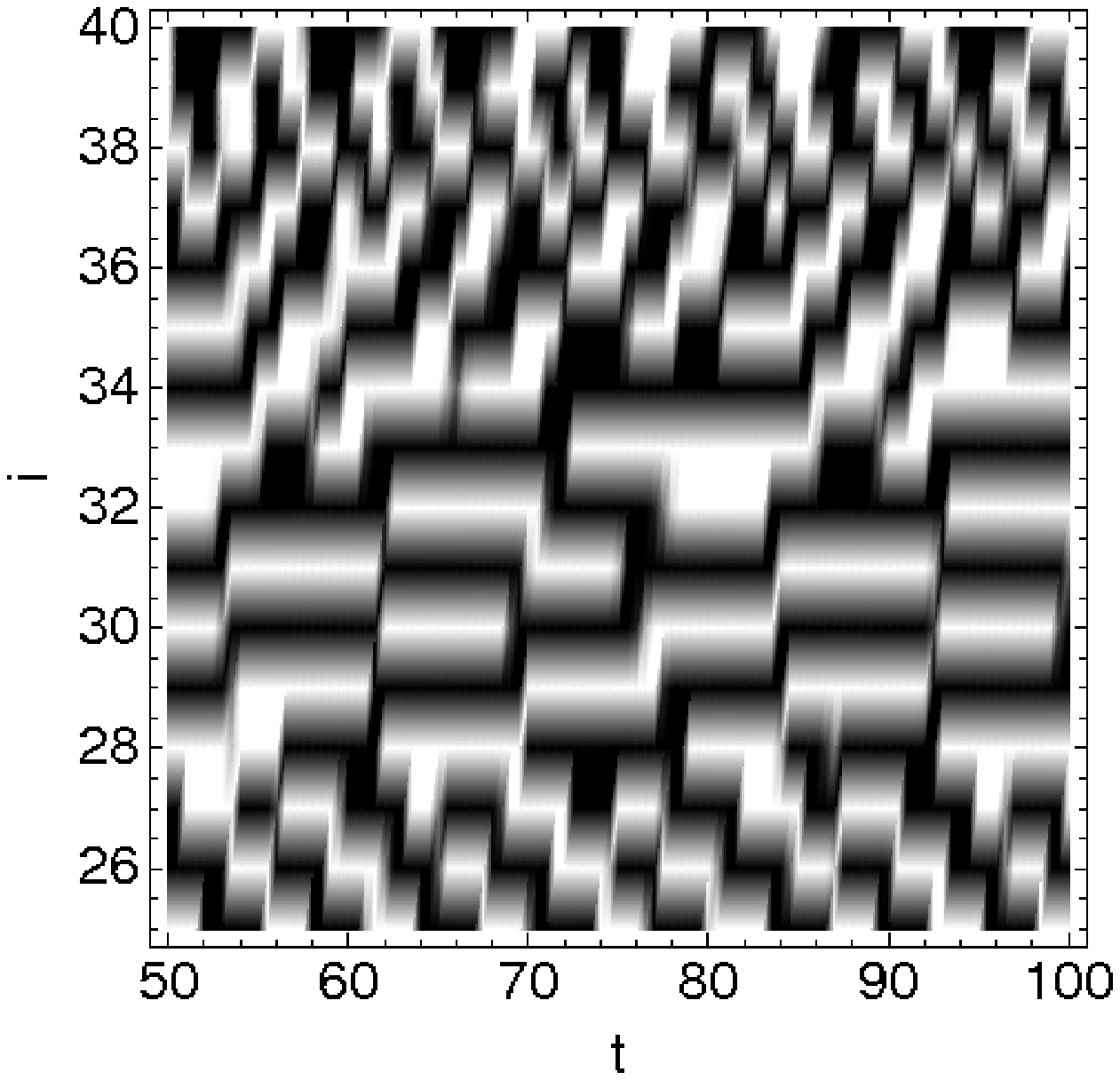}\\
	\end{tabular}
\end{center}
\caption{Phase difference plots(left) and space-time plots(right) at four different coupling strengths for the ring geometry. The space-time plots show a selection of $15$ oscillators  taken from a $64$ oscillator system. Time is along the horizontal  axes and space is along the vertical axes. }
\label{fig-spacetime}
\end{figure}


\begin{figure*}
\begin{tabular}{cc}
	\includegraphics[width=\columnwidth]{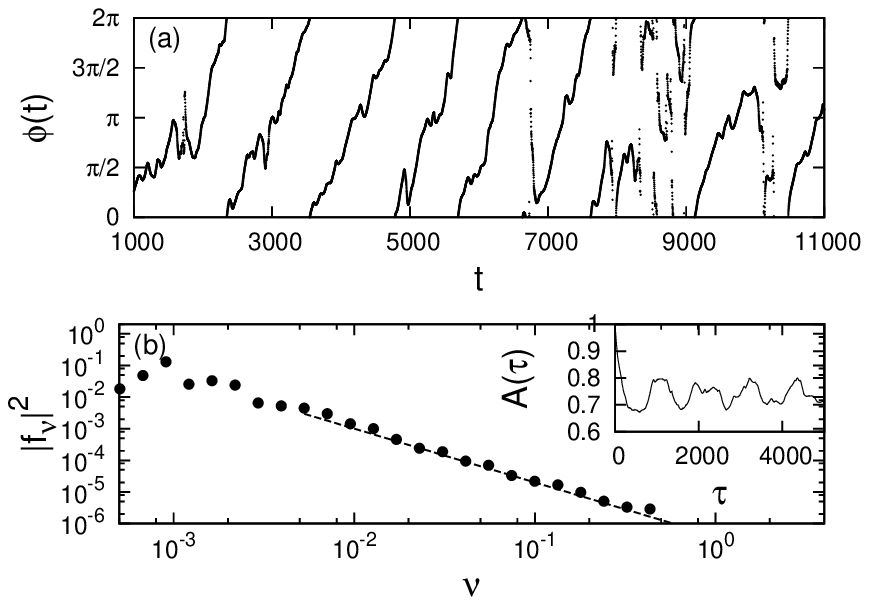}&
\includegraphics[width=\columnwidth]{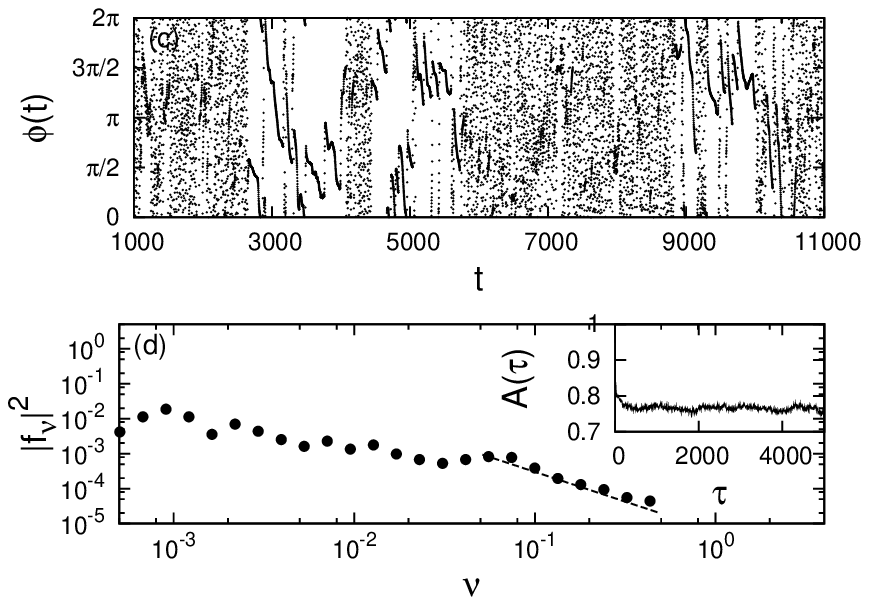}\\
\end{tabular}
	\caption{  (a),(c) Time-series of the phase $\phi(t)$ of a typical oscillator in the interior and at the boundary of a domain respectively, at coupling strength  $K=3.0$. (b),(d) Power spectrum $|f_{\nu}|^2$ calculated from the above time-series. The power spectrum  $|f_{\nu}|^2$ shows a power-law decay with an exponent $\sim 1.7$. 
The inset to the plots (b),(d) show the corresponding auto-correlation function $A(\tau)$
The solutions were sampled after an initial transient $t_0=1000$ over a time interval  $T=10000$. 
	\label{powspec}}
\end{figure*}
\subsection{Frequency Entrainment}
When the coupling between the oscillators is turned off, each of the oscillators in the system evolves in time according to its own frequency, $\omega_i$. As we switch on and increase the coupling, we observe the oscillators to form local frequency entrained clusters. As the coupling is increased further, the clusters merge with one another until eventually, at some critical value $K_c$, all of the oscillators are entrained with a common frequency. This clustering process is illustrated in Fig. \ref{fig-synchtree}.  By carrying out our study at different values of the randomness, $\sigma$, we have directly tested that the parameter controlling the synchronization is $K/\sigma$: increasing the variance of the $\omega_i$  distribution has the same effect as reducing the coupling.   As Fig. \ref{fig-critcoupling} shows, the critical coupling strength obtained from frequency synchronization increases as the number of oscillators is increased, and there is no finite-$K$ transition in the thermodynamic limit. We note that linear chain requires a larger coupling strength to obtain frequency entrainment than a ring system with the same number of oscillators and $K_c$ shows no systematic dependence on odd/even values of $N$ for either geometry.  Clustering behavior in the frequency synchronization of a ring of oscillators has been observed earlier in a locally coupled Kuramoto model with attractive coupling\cite{Nashar2009}. 
In general, we find that even though the character of frequency synchronization in  this repulsively coupled system is qualitatively similar to that seen in attractively coupled systems, the phase patterns obtained can be significantly different.

Fig. \ref{fig-synchtree} does not give any information about how the oscillators in a given cluster are distributed in space.  Fig. \ref{fig-spacetime} provides insight into how the clustering of frequencies occurs spatially.   The figure shows four space-time plots over a range of coupling and a corresponding plot of the phase difference as a function of position on the lattice. At low coupling both the frequencies and phases are disordered.  There is, however, a structure visible in the space-time plots even at $K/\sigma=1$ showing that the oscillators in the same frequency cluster are spatially correlated.  As the coupling is increased to $K/\sigma = 10$ we can see that all the oscillators evolve with approximately the same frequency, but maintain some disorder in the phase relationships. It is not until a coupling strength two orders of magnitude larger that we see perfect phase ordering at this system size. The patterns  that we obtain from our model is qualitatively similar to that observed in experiments on BZ droplets \cite{Toiya:2008fk,Toiya:2010kx}.

\section{Results in two dimensions \label{sec3}}

In the two dimensional experimental setup \cite{Toiya:2010kx}, the droplets containing the BZ reactants reorganize into an hexagonal array.  We model this situation by placing the Kuramoto oscillators on a triangular lattice, each oscillator being repulsively coupled to its six nearest neighbors. As was shown in the one-dimensional model, the phases of the oscillators tend to align $\pi$ out of phase with their neighbors, when they are repulsively coupled. The locally-preferred order of $\pi$ phase difference between nearest-neighbor oscillators is not globally realizable on the triangular lattice which has fundamental loops containing three oscillators that are mutual nearest neighbors; a situation analogous to the three-oscillator ring discussed in section \ref{sec2}.  As we show below, this frustration leads to (a) much richer dynamics in comparison to attractive coupling, (b) defects in the phase order, and (c) a phase-disordered yet frequency synchronized state.

\begin{figure}[!b]
	\includegraphics[width=\columnwidth]{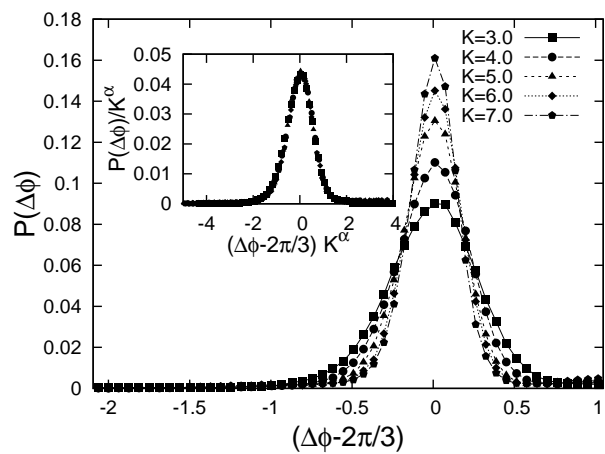}
	\caption{Distribution of  the magnitude of phase differences  between two adjoining oscillators on a row $P(\Delta\phi)$, plotted for various coupling strengths $K$. The distribution is a gaussian which peaks around a phase difference of $2\pi/3$. The inset shows the rescaled distribution $P(\Delta\phi) K^{-\alpha}$ plotted against $(\Delta\phi-2\pi/3)K^{\alpha}$. $\alpha=2/3$. \label{phasedist}}
\end{figure}

\begin{figure*}
	\begin{tabular}{ccc}
		\includegraphics[height=2.4in]{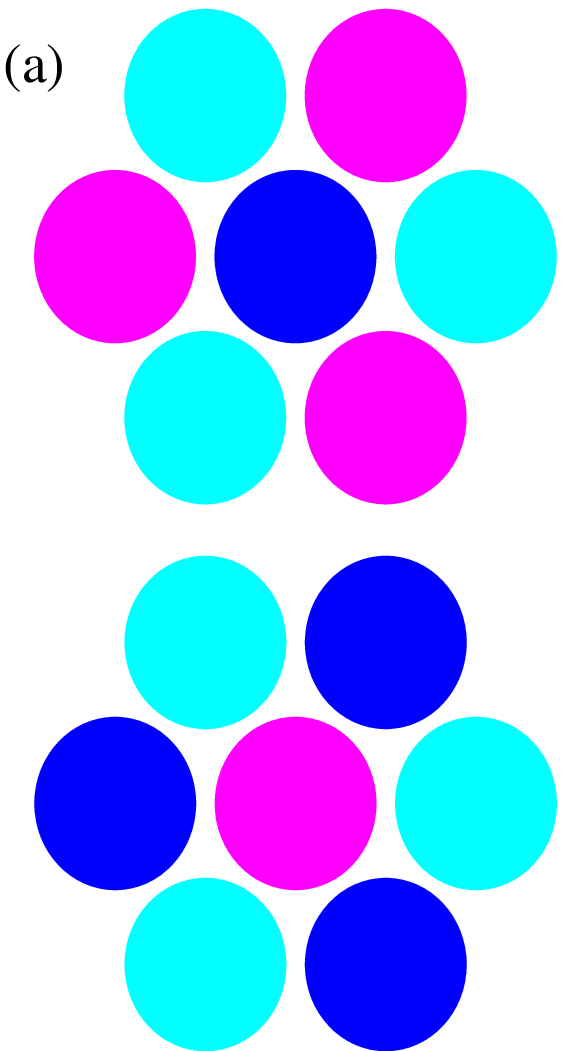}&
\includegraphics[height=2.5in]{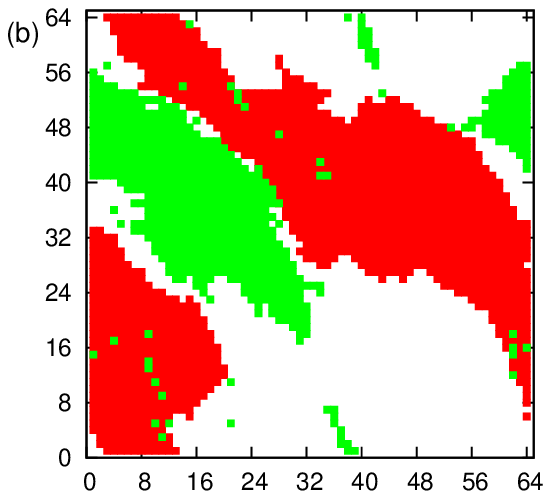}&
		\includegraphics[height=2.55in]{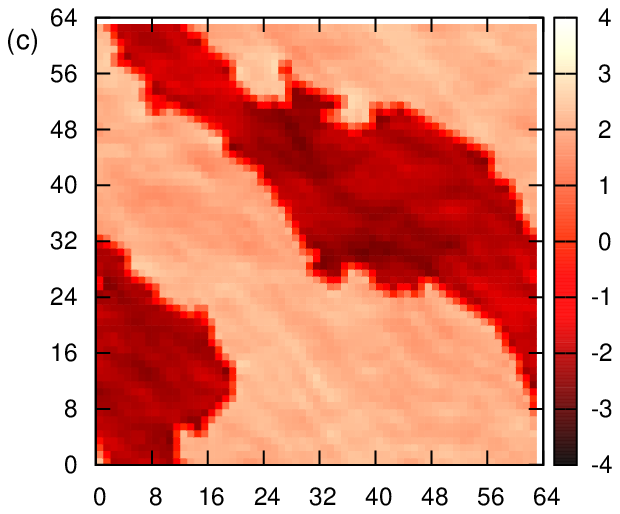}\\
	\end{tabular}
	\caption{(Color online)  (a) Graphic illustrating the different helicities seen on the triangular lattice. (b) Domains of opposite helicities represented by the red  (darker) and white regions seen for coupling strength $K=3.0$. The green (lighter) regions indicate the frequency entrained oscillators. (c) Colorscale map  of the phase differences between two oscillators on a row. The dark and lighter regions  represent location of oscillators with a phase difference of $-2\pi/3$ and $2\pi/3$  respectively. \label{graphic}}
\end{figure*}

 The rich and spatially heterogenous dynamics in the case of repulsively coupled oscillators is evident in Fig.  \ref{powspec}(a),(c), in which the time-series of 
the phases  $\phi(t)$ of two oscillators at different locations in the lattice, at a coupling strength $K=3.0$, is shown (As will be discussed later, these oscillators are located in the interior and at the boundary of a phase-locked domain respectively).
The two oscillators exhibit widely different dynamics. 
While the first oscillator  shows relatively monotonic behavior, the latter oscillator exhibits intermittent dynamics in which the oscillator follows a monotonic trajectory in between bursts of chaotic dynamics.
The power spectrum $|f_{\nu}|^2$ of these oscillators obtained by taking a Fourier transform of the time series $\phi(t)$  ($f_{\nu}=\int dt~ \phi(t) \exp(i\nu t)$),
  exhibits a power-law decay $|f_{\nu}|^2 \sim \nu^{-\zeta}$ with an exponent $\zeta\sim1.7$  (Fig. \ref{powspec}(b),(d)). This spectrum indicates the existence of a multitude of time-scales in the phase dynamics of the repulsively coupled oscillators.  The exponent $\zeta$ varies between $1.7-2.0$ for different oscillators, and is typical of the spatial heterogeneity seen in the lattice.
The auto-correlation function $A(\tau)$ calculated by taking an inverse Fourier transform of the power-spectrum $|f_{\nu}|^2$, $A(\tau)=\int d\nu |f_{\nu}|^2 \exp(-i\nu\tau)$, shown in the insets to Fig. \ref{powspec}(b),(d), shows a slow relaxation.  

 Non-exponential relaxations and multiple time scales are a  hallmark of glassy systems \cite{spinglassbook}, and reflect the presence of many metastable states, and a complex free-energy landscape.  In the coupled oscillators system, the multiple time scales presumably arise from a complex attractor landscape.  Within the phase-coupled model, we therefore expect to see glassy behavior, and the correlation functions discussed below, indeed indicate frozen, disordered phase patterns.

At weak coupling strengths $K$, the phases of the oscillators are mainly disordered, though a few neighboring oscillators show a tendency to oscillate $\pi$ out of phase with each other.  As the coupling strength is increased above  $K>K_p \simeq 1.5$,  the oscillators relax into a state in which each oscillator is $2\pi/3$ out of phase with its neighbor. 
This resembles the $2\pi/3$ state seen in the BZ micro-oscillators setup \cite{Toiya:2010kx}. 
The distribution of the magnitude of the phase differences  between neighboring oscillators $P(\Delta\phi)$, shows a peak at $2\pi/3$ (Fig. \ref{phasedist}). The distributions at different coupling strengths can be collapsed onto a single curve, when the phase differences are rescaled as given below.
\begin{equation}
P(\Delta\phi,K)= K^{-\alpha} f((\Delta\phi-2\pi/3) K^{\alpha})
\end{equation}
The inset to Fig. \ref{phasedist}, in which the rescaled distribution of phase differences is plotted for different coupling strengths, illustrates the scaling behavior. 
The form of the scaling function $f((\Delta\phi-2\pi/3) K^{\alpha})$ is seen to be gaussian. Hence, the variance of the distributions is proportional to $1/K^{2\alpha}$, where $\alpha=2/3$. This suggests that as the coupling strength approaches infinity, the distribution approaches a $\delta-$function centered at $2\pi/3$. Hence, oscillators prefer to relax into a state with a phase difference of $2\pi/3$.

Interestingly, however, at coupling strengths  $K>K_p$, domains with opposite helicities form in the lattice. In each domain, 
the phases of any three neighboring oscillators vary 
continuously in either  clockwise or an anti-clockwise direction. 
The two opposite helicities have been illustrated by a graphic in Fig. \ref{graphic}(a). The two different domains on the lattice, have been represented by the red (darker) and white regions in Fig. \ref{graphic}(b). 
The phase differences between two neighbors in a row in the two domains could be either $2\pi/3$ or $-2\pi/3$ depending on the helicity.
This is shown in Fig. \ref{graphic}(c), where the phase differences in the lattice has been plotted. The  darker regions represent location of oscillators with a phase difference of $-2\pi/3$ between the neighbors. 
At the boundaries of these domains, the phases of the neighboring oscillators are seen to be aligned anti-parallel with each other. So, in a lattice,  
a non-zero fraction of neighboring oscillators are $\pi$ out of phase with each other, and the  fraction of $\pi$ out-of-phase neighbors is proportional to the fraction of sites belonging to the domain boundaries. 
In the interior of a domain, the helicity of the phases are frozen.
However, the oscillators at the 
domain boundaries can fluctuate between either of the helicities, since they exist in an anti-phase state. 
 This is evident in the time-series of the phase $\phi(t)$ of a boundary 
oscillator plotted in Fig \ref{powspec}(c), in which the phase 
exhibits intermittent dynamics
 when the domain boundary shifts positions in the lattice.
These domain boundaries then act as defects in the lattice. 
At weak coupling strengths ($K_p<K<3.5$),  the domains show a tendency to coarsen with time. In other words, domains with one of the helicities grow resulting in a shrinkage in the size of domains with the opposite helicity. The selection of the dominant helicity is unbiased in a set of random initial conditions. As the coupling strength is increased further $(K>3.5)$, the domains freeze after a very short transient, and do not change with time. 

In analogy with Ising spin systems \cite{XYmodel_landau}, we  define a helicity order parameter  that characterizes the domains on the lattice. We measure the phase relation between any three neighboring sites $1,2,3$ as shown in the graphic in Fig.  \ref{graphic}(a), and check the helicity of this triplet. We associate a helicity index $\rho(R)=+1$ or $-1$ with site $1$  if the helicity is clockwise or anti-clockwise respectively. 
If $N_+$ and $N_-$ are the total number of sites in each domain, we define the helicity order parameter $m$ as 
$m=|N_+-N_-|/N$.  The dependence of the helicity order parameter $m$ on the coupling strength is shown in Fig. \ref{hel_op}.
 The helicity order parameter $m$ is close to zero when the phases are disordered. As the coupling strength $K$ is increased, domains with opposite helicities are formed. Due to the coarsening of the domains and the subsequent domination of one of the helicities on the lattice, the helicity order parameter $m$ approaches $1$. On further increasing the coupling strength, frozen domains with finite sizes are seen, and 
an order parameter $m<1$ is obtained.
A clearer signature of the onset of the domain 
formation is seen in the fluctuation of the helicity order parameter. When the phases are disordered, viz $K<K_p$, the  fluctuation in the order parameter $\langle\Delta m^2\rangle=\langle(m-\bar{m})^2\rangle$ is negligible. When distinct domains start forming in the lattice, as the coupling strength is increased beyond $K_p$, the fluctuation in the order parameter $\langle\Delta m^2\rangle$ shows a distinct increase, as shown in Fig. \ref{hel_op}.

\begin{figure}
	\includegraphics[width=\columnwidth]{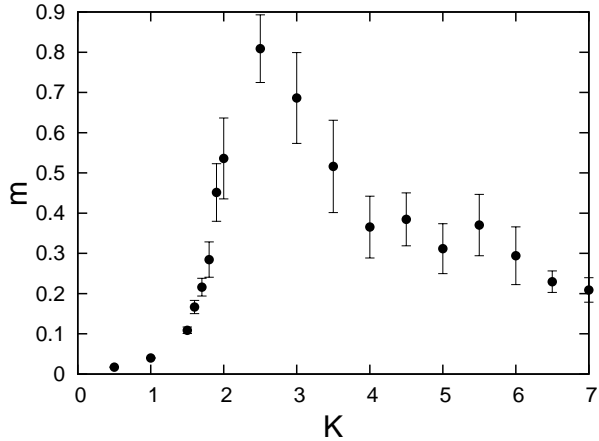}
	\caption{Helicity order parameter $m$ plotted as a function of the coupling strength $K$. The error bars represent the fluctuations in the order parameter $\langle\Delta m^2\rangle$. The data has been collected after  a transient time $t_0=5000$. \label{hel_op} }
\end{figure}

\begin{figure}
	\begin{tabular}{c}
		\includegraphics[width=\columnwidth]{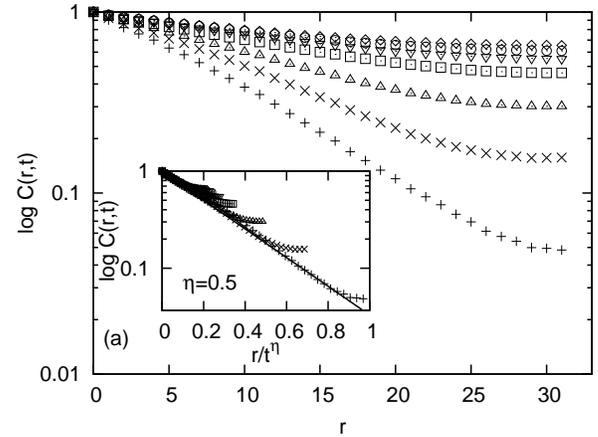}\\
		\includegraphics[width=\columnwidth]{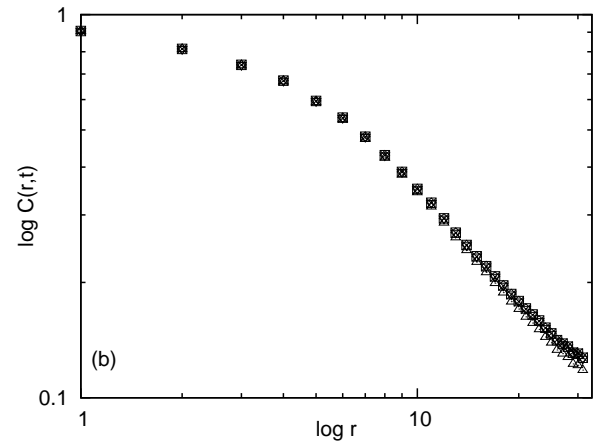}\\
	\end{tabular}
\caption{ (a) Spatial correlation function $C(r,t)$ plotted at coupling strength $K=3.0$ and at times $t=1024(+), 2048(\times), 4096(\triangle), 8192(\Box), 12288(\nabla), 16384(\bigcirc), 20480(\Diamond)$. The inset shows the correlation function $C(r,t)$ plotted against the scaled variable $r/t^{\eta}$, where $\eta=0.5$. The dashed line represents an exponential fit to the data, $\exp(-3.4 x)$. (b) Spatial correlation function $C(r,t)$ plotted for coupling strength $K=5.0$  at times $t=4096(\triangle), 8192(\times), 12288(\nabla), 16384(\Box), 20480(\Diamond)$. \label{corrfn}}
\end{figure}

\begin{figure}
\includegraphics[width=\columnwidth]{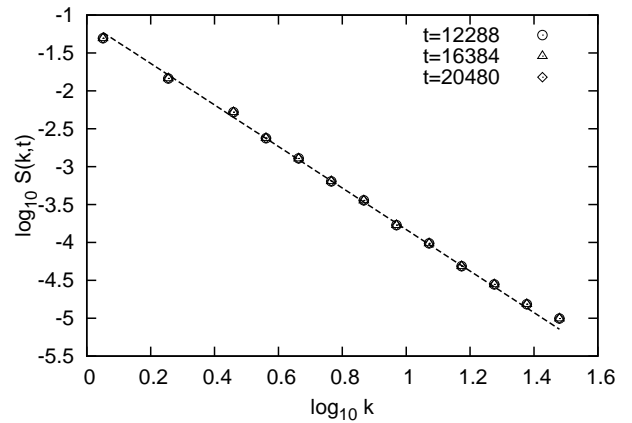}
\caption{Structure factor plotted for coupling strength  $K=5.0$. Dashed line shows a power-law fit to the data with an exponent $-2.74\pm 0.04$.\label{strfac} }
\end{figure}

  Signatures of the dynamics of domains can be detected in the time-dependent  spatial correlation function $C(r,t)$, which is defined as  
$C(r,t)=\int dR \rho(R,t)\rho(R+r,t)$, where $\rho(R,t)$ is the helicity index $+1,-1$ associated with the site at position $R$.
At weak coupling strengths ($K_p<K<3.5$), where the domains coarsen with time, the correlation function $C(r,t)$ shows an increasing correlation length as time progresses (Fig. \ref{corrfn}(a)). The domain size $L(t)$ scales with time $t$ as $L(t)\sim t^{\eta}$, where $\eta=0.5$.This is seen in the inset to Fig. \ref{corrfn}(a), in which the correlation function obeys the scaling form $C(r,t) \sim f(r/t^{\eta})$. The scaling function $f(x)$ is seen to be an exponential of the form $f(x)=\exp(-ax)$. An exponent of $\eta=0.5$  is usually seen in the case of non-conserved scalar fields, where the domain growth is surface tension-driven \cite{bray}.  At larger coupling strengths,  the correlation function does not evolve with time after a transient period, concurrent with the frozen domains observed at these parameter values (Fig. \ref{corrfn}(b)).

The structure factor $S(k,t)$, defined as the fourier transform of the spatial correlation function $C(r,t)$ corroborates the formation of distinct domains in the lattice.  As shown in Fig. \ref{strfac}, the structure factor $S(k,t)$ decays as $S(k,t)\sim k^{-\theta}, \theta=2.74\pm0.04 \simeq 3$ for large wavevectors $k$. This is in accordance with Porod's law wherein the structure factor decays as $S(k,t)\sim 1/k^{d+1}$ whenever there are well-defined domain boundaries \cite{porod}.

 A  possible reason for the coarsening of domains at weak coupling strengths can be inferred from  the collective behavior of the frequencies of oscillators. We define an effective frequency, $\omega_{eff}$ for each oscillator 
as 
\begin{equation}
\omega_{eff}=(\phi_i(t_0+\tau)-\phi_i(t_0))/\tau																																				
\end{equation}
Here, the frequencies have been averaged over the interval $\tau=5000$ after discarding data for a  transient time $t_0=1000$. 
The inset to Fig. \ref{freqop} shows a bifurcation diagram, in which all the frequencies of the oscillators have been plotted as a function of the coupling strength $K$. 
At weak coupling strengths, the frequencies of the oscillators are not entrained, and a broad  distribution of the oscillator frequencies is obtained. 
In this interval of coupling strength, the phases of the oscillators also evolve with different frequencies, hence we see the fluctuating domain boundaries, and growing domains. In this regime, the frequency entrained oscillators are seen in the interior of the domains, as shown in Fig. \ref{graphic}(b).

\begin{figure}
	\includegraphics[width=\columnwidth]{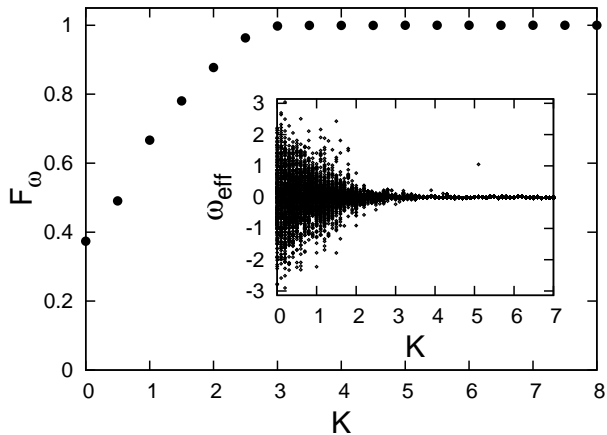}
	\caption{Frequency order parameter $|F_{\omega}|$ plotted as a function of the coupling strength $K$. The inset shows a bifurcation diagram, in which the time-averaged frequencies of the oscillators $\omega_{eff}$ have been plotted as a function of the coupling strength $K$. \label{freqop}}
\end{figure}

At larger coupling strengths, the oscillator frequencies are well entrained, and hence the phase patterns on the lattice are frozen.  A quantitative measure of frequency entrainment is given by the frequency order parameter $F_{\omega}$ \cite{Hong:2005ss}, which is defined as 
\begin{equation}
F_{\omega}=<N_n/N>
\end{equation}
where $N_n$ is the maximum number  of oscillators with identical frequencies, and $N$ is the total number of oscillators. When all the oscillators are frequency entrained, the order parameter approaches $1$. The frequency order parameter $F_{\omega}$ has been plotted as a function of coupling strength $K$ in Fig. \ref{freqop}. 
We see that the coupling strength $K=3.5$ at which complete frequency entrainment is seen, or $F_{\omega}=1$, coincides with the occurrence of frozen domains. Hence, frequency entrainment is responsible for the frozen domains in this lattice at larger values of the coupling strengths.   The freezing of the domains is consistent with the form of the power spectra of phases shown in Fig. \ref{powspec}, which indicates a broad distribution of time scales in the system.  The relative phases between neighboring oscillators can, therefore, be frozen in a disordered state while the whole system oscillates with a common frequency.  The fundamental reason for the appearance of a broad time-spectrum is not clear from our studies.  We, however, expect that this feature is related to the tendency of neighboring oscillators to prefer a phase difference of $\pi$, thus freezing in domain boundaries where this configuration is realizable.  We are continuing to investigate the slow dynamics in this model.

\section{Discussion \label{sec5}}

In this paper, we studied locally coupled Kuramoto oscillators with repulsive coupling in one and two dimensions. In comparison to the attractively coupled system, the system studied shows much richer dynamics, with hints of multiple time-scales, and a complex attractor landscape, which strongly depends on the underlying geometry of the lattice. 

In one dimension we showed that while the linear chain has one stable phase configuration, when periodic boundary conditions are introduced a range of attractors become available to the system which depend upon the number of oscillators. The spatial patterns obtained qualitatively match with 
anti-phase patterns seen in the BZ micro-droplets experimental setup.
Additionally, we showed that even though the frequencies were entrained, the phase patterns were still disordered. Hence, the onset of frequency entrainment occurs at a lower coupling than the phase ordering. 
In two dimensions,  we showed the existence of phase patterns similar to the $2\pi/3$ state seen in the BZ micro-oscillator system. We showed the existence of  domains with clockwise and anti-clockwise helicities in the same lattice. These   domains 
showed coarsening behavior at weak coupling strengths, where a growing length scale could be detected that reached system size at large times. As the coupling strength was increased, we found that these domains freeze, such that the phase pattern does not change in time. This was attributed to frequency entrainment at large coupling strengths, which ensured that the frozen  phase patterns on the lattice oscillate with a common frequency. 
Mapping discrete dynamical systems to statistical mechanics models gives new insights into the behavior of these systems
\cite{zahera,zaherapla}.
 The repulsively coupled Kuramoto oscillators serve as a good paradigm  in which techniques and ideas related to statistical mechanics can be applied to  an inherently dynamical system.

Since Kuramoto phase oscillator formalism excludes the discussion of amplitude variations, we do not hope to see the $\pi-$S state seen in the experimental system in this model. However, we are 
looking at an 
extension of this model, which allows the oscillators to "switch" on or off based on its phase environment.  

The authors acknowledge partial support of this research by the donors of the American Chemical Society Petroleum Research Fund, and by Brandeis NSF-MRSEC.   MG has also been supported by a NSF-IGERT fellowship.   We acknowledge many useful discussions with Irv Epstein, Seth Fraden, Ning Li, Hector Gonzalez-Ochoa, Mitch Mailman and Dapeng Bi.

\bibliography{repcoup_kuramoto}
\end{document}